\documentclass{raa}           
\usepackage{graphicx,times}
\usepackage{natbib}
\usepackage{amssymb,amsmath}
\bibpunct{(}{)}{;}{a}{}{,}

\usepackage[a4paper=true,dvipdfm=true,pagebackref=true]{hyperref}
\hypersetup{pdftitle = The title of my PDF, pdfauthor = My name, pdfsubject= The subject, pdfkeywords = keyword1 keyword2 keyword3} 
\hypersetup{colorlinks = true, linkcolor = green, anchorcolor = red, citecolor = blue, filecolor = red, pagecolor = red, urlcolor = red}

\begin{document}

   \title{Near-Infrared Polarimetry of the GG Tauri A Binary System $^*$
\footnotetext{\small $*$ Based on data collected at Subaru Telescope, which is operated by the National Astronomical Observatory of Japan.}
}

 \volnopage{ {\bf 2014} Vol.\ {\bf X} No. {\bf XX}, 000--000}
   \setcounter{page}{1}

   \author{Yoichi Itoh\inst{1}, Yumiko Oasa\inst{2},
           Tomoyuki Kudo\inst{3}, 
Nobuhiko Kusakabe\inst{4}, 
Jun Hashimoto\inst{5}, 
Lyu Abe\inst{6}, 
Wolfgang Brandner\inst{7}, 
Timothy D. Brandt\inst{8}, 
Joseph C. Carson\inst{9}, 
Sebastian Egner\inst{3}, 
Markus Feldt\inst{9}, 
Carol A. Grady\inst{10,11,12}, 
Olivier Guyon\inst{3}, 
Yutaka Hayano\inst{3}, 
Masahiko Hayashi\inst{4}, 
Saeko S. Hayashi\inst{3},
Thomas Henning\inst{7}, 
Klaus W. Hodapp\inst{13}, 
Miki Ishii\inst{3}, 
Masanori Iye\inst{4}, 
Markus Janson\inst{8}, 
Ryo Kandori\inst{4}, 
Gillian R. Knapp\inst{8}, 
Masayuki Kuzuhara\inst{14},
Jungmi Kwon\inst{4}, 
Taro Matsuo\inst{15}, 
Michael W. McElwain\inst{10},
Shoken Miyama\inst{16}, 
Jun-Ichi Morino\inst{4}, 
Amaya Moro-Martin\inst{8,17}, 
Tetsuo Nishimura\inst{3}, 
Tae-Soo Pyo\inst{3},
Eugene Serabyn\inst{18}, 
Takuya Suenaga\inst{4,19}, 
Hiroshi Suto\inst{4}, 
Ryuji Suzuki\inst{4}, 
Yasuhiro H. Takahashi\inst{20,4}, 
Naruhisa Takato\inst{3}, 
Hiroshi Terada\inst{3}, 
Christian Thalmann\inst{21}, 
Daigo Tomono\inst{3}, 
Edwin L. Turner\inst{8,22},
Makoto Watanabe\inst{23}, 
John Wisniewski\inst{5}, 
Toru Yamada\inst{24}, 
Satoshi, Mayama\inst{25},
Thayne Currie\inst{26},
Hideki Takami\inst{4}, 
Tomonori Usuda\inst{4},
Motohide Tamura\inst{20,4}
}
%% Here is an example of three authors come from different institutes.
%% For single author or all the authors from an institute, use "\inst{}" only

%% Please give the E-mail address of the author, to whom future correspondence and
%% offprint requests will be sent.
\institute{Nishi-Harima Astronomical Observatory, Center for Astronomy, 
             University of Hyogo, 
             407-2, Nishigaichi, Sayo, Hyogo 679-5313, Japan
             {\it yitoh@nhao.jp} \\
\and
Faculty of Education, Saitama University,
             255 Shimo-Okubo, Sakura, Saitama, Saitama 338-8570, Japan\\
\and
Subaru Telescope, National Astronomical Observatory of Japan, 650 North A'ohoku Place, Hilo, HI96720, USA\\
\and
National Astronomical Observatory of Japan, 2-21-1, Osawa, Mitaka, Tokyo, 181-8588, Japan\\
\and
H.L. Dodge Department of Physics \& Astronomy, University of Oklahoma, 440 W Brooks St. Norman, OK 73019, USA\\
\and
Laboratoire Lagrange (UMR 7293), Universit\'e de Nice-Sophia Antipolis, CNRS, Observatoire de la Cote d'Azur, 28 avenue Valrose, 06108 Nice Cedex 2, France\\
\and
Max Planck Institute for Astronomy, K\"onigstuhl 17, 69117 Heidelberg, Germany\\
\and
Department of Astrophysical Science, Princeton University, Peyton Hall, Ivy Lane, Princeton, NJ08544, USA\\
\and
Department of Physics and Astronomy, College of Charleston, 58 Coming St., Charleston, SC 29424, USA\\
\and
Exoplanets and Stellar Astrophysics Laboratory, Code 667, Goddard Space Flight Center, Greenbelt, MD 20771, USA\\
\and
Eureka Scientific, 2452 Delmer, Suite 100, Oakland CA96002, USA\\
\and
Goddard Center for Astrobiology\\
\and
Institute for Astronomy, University of Hawaii, 640 N. A‘ohoku Place, Hilo, HI 96720, USA\\
\and
Department of Earth and Planetary Sciences, Tokyo Institute of Technology, Ookayama, Meguro-ku, Tokyo 152-8551, Japan\\
\and
Department of Astronomy, Kyoto University, Kitashirakawa-Oiwake-cho, Sakyo-ku, Kyoto, Kyoto 606-8502, Japan\\
\and
Hiroshima University, 1-3-2, Kagamiyama, Higashihiroshima, Hiroshima 739-8511, Japan\\
\and
Department of Astrophysics, CAB-CSIC/INTA, 28850 Torrej\'on de Ardoz, Madrid, Spain\\
\and
Jet Propulsion Laboratory, California Institute of Technology, Pasadena, CA, 171-113, USA\\
\and
Department of Astronomical Science, The Graduate University for Advanced Studies, 2-21-1, Osawa, Mitaka, Tokyo, 181-8588, Japan\\
\and
Department of Astronomy, The University of Tokyo, 7-3-1, Hongo, Bunkyo-ku, Tokyo, 113-0033, Japan\\
\and
Astronomical Institute "Anton Pannekoek", University of Amsterdam, Postbus 94249, 1090 GE, Amsterdam, The Netherlands\\
\and
Kavli Institute for Physics and Mathematics of the Universe, The University of Tokyo, 5-1-5, Kashiwanoha, Kashiwa, Chiba 277-8568, Japan\\
\and
Department of Cosmosciences, Hokkaido University, Kita-ku, Sapporo, Hokkaido 060-0810, Japan\\
\and
Astronomical Institute, Tohoku University, Aoba-ku, Sendai, Miyagi 980-8578, Japan\\
\and
The Center for the Promotion of Integrated Sciences, The Graduate University for Advanced Studies (SOKENDAI), Shonan International Village, Hayama-cho, Miura-gun, Kanagawa 240-0193, Japan\\
\and
Department of Astronomy \& Astrophysics, University of Toronto, 50 George St., Toronto, Ontario, M5S 3H4, Canada\\
\vs \no
   {\small Received 2013 November 29; accepted 2014 May 5}
}

\abstract{
A high angular resolution near-infrared polarized-intensity image of the GG
Tau A binary system was obtained with the Subaru Telescope.
The image shows the circumbinary disk scattering the light from the central 
binary.
The azimuthal profile of the polarized intensity of the circumbinary disk 
is roughly reproduced by a simple disk model
with the Henyey-Greenstein function and the Rayleigh function, 
indicating small dust grains
at the surface of the disk.
Combined with a previous observation of the circumbinary disk,
our image indicates that the gap structure in the circumbinary disk orbits
anti-clockwise,
while material in the disk orbit clockwise.
We propose
a shadow of material located between the central binary and the circumbinary
disk.
The separations and position angles of the stellar components of the
binary in the past 20 years are consistent with the binary orbit with
$a = 33.4$ AU and $e = 0.34$.
}

   \authorrunning{Y. Itoh et al.}            %author_head in even pages
   \titlerunning{Polarimetry of GG Tau}  % title_head in odd pages
   \maketitle

%________________________________________________ sections below
%
\section{Introduction}           %% first-level sections will be auto-capitalized
\label{sect:intro}

Proto-planetary disks are common structures
around classical T Tauri stars.
A number of disks have been investigated at various wavelengths. 
However, many observations have focused on the disks around 
single stars, while more than half of T Tauri stars are binaries 
(\citealt{Ghez1993}; \citealt{Leinert}). 
Disks around binaries are expected to have forms different from those
around single stars.
\cite{Artymowicz} indicated two kinds of disks around a binary system:
a circumstellar disk associated with each star and 
a ring-shaped circumbinary disk around the binary system. 
A cavity exists between the circumbinary disk and the central binary. 
Several circumbinary disks have been spatially resolved at the near-infrared
wavelengths (e.g. UY Aur; \citealt{Hioki}, FS Tau; \citealt{Hioki11}).

GG Tau ($d\sim$ 140 pc) is a well-studied young multiple system.
The system has two binaries: GG Tau Aa/Ab and GG Tau Ba/Bb.
The GG Tau A binary (hereafter GG Tau) is especially interesting, because
its circumbinary disk has been spatially resolved at
the millimeter wavelengths (\citealt{Guilloteau}), at the near-infrared
(\citealt{Roddier}; \citealt{Itoh}), and at
the optical (\citealt{Krist}).
The large-scale structure of the circumbinary disk is well characterized by 
an annulus with an inner radius of 190 AU.
The disk is inclined by $\sim$37\degr with the northern edge 
nearest to us.
The kinematics of the disk are consistent with clockwise Keplerian 
rotation (\citealt{Guilloteau}).
\cite{Duchene} observed GG Tau at the $L'$-band
($\lambda=3.8 \mu$m).
Comparing this with shorter wavelength images,
they proposed a stratified structure for the circumbinary disk, 
in which large dust
grains are present near the disk midplane.
They suggested vertical dust settling and grain growth
in the dense part of the disk.

Polarimetric observations also yield insight into dust properties
and disk structures.
\cite{Tanii} carried out near-infrared polarimetry for UX Tau.
The circumstellar disk of UX Tau shows large variety
in the polarization degree, from 1.6 \% up to 66 \%.
They attributed this large variation to non-spherical large dust
grains in the circumstellar disk.
This observation confirmed dust growth in a circumstellar
disk.
On the other hand, several observations suggest
small dust grains in proto-planetary disks.
\cite{Silber} carried out polarimetric observations of
GG Tau at 1 $\mu$m.
They found that the circumbinary disk is strongly
polarized, up to $\sim 50$ \%, which is indicative of
Rayleigh-like scattering from sub-micron dust grains. 

We present the results of near-infrared polarimetry
of GG Tau. Combining a coronagraph with an adaptive optics system,
we obtained a high spatial resolution polarized-intensity
image of GG Tau.
The observations and the data-reduction procedure
are described in section 2.
In section 3, we discuss 
the circumbinary structures around GG Tau and the orbital motion of the GG Tau 
binary.

\section{Observations and Data Reduction}

Near-infrared $H$-band (1.6 $\mu$m) polarimetric
imaging observations of GG Tau A were carried out on
September 4, 2011 with the High Contrast Instrument for the
Subaru next-generation Adaptive Optics (HiCIAO) and the
adaptive optics system, AO188, mounted on the Nasmyth platform
of the Subaru Telescope.
The observations were conducted as part of the SEEDS survey.
We employed the Polarization Differential Imaging (PDI) mode.
In this mode,
the Wollaston prism installed in HiCIAO divides the incident light into
two linearly polarized components, which are perpendicular to
each other and imaged simultaneously on the detector.
Each image has 1024 $\times$ 2048 pixels with a field of view of 
9\farcs75 $\times$ 20\farcs09 and a pixel scale of 9.521 mas
pixel$^{-1}$ in the east-west direction and 9.811 mas pixel$^{-1}$
in the north-south direction.
In the PDI mode, when the half-wave plate is set at an offset angle
of 0\degr, 45\degr, 22\fdg5, and 67\fdg5, we obtained
polarimetric images with the polarization direction at 0\degr~and 90\degr,
90\degr~and 0\degr, 45\degr~and 135\degr, 
and 135\degr~and 45\degr components, respectively.
The full width at half maximum of the point spread function (PSF)
was 0\farcs11.
We used a coronagraphic mask with 0\farcs6 diameter to
suppress the brightness of GG Tau Aa/Ab.
We obtained 64 frames with an exposure time of 30 s for each.
We also took short exposure frames without the coronagraphic mask.
For a PSF reference star, 
we took SAO 76661 after the GG Tau observations
with the same instrumental configuration except for use of the 0\farcs3
diameter coronagraphic mask. 
We used an ND filter in the AO system to match the $R$-band
magnitudes between GG Tau and SAO 76661 where wavefronts were sensed.
Twelve frames were taken with an exposure time of 30 s.

The Image Reduction and Analysis Facility (IRAF\footnote{IRAF is distributed by National Optical Astronomy Observatory, which is operated by the Association of Universities for Research in Astronomy, Inc., under cooperative agreement with the National Science Foundation.}) software
was used for data reduction.
We followed reduction procedures given by \cite{Tanii}.
All HiCIAO frames have artifacts of horizontal stripes and vertical bandings.
These patterns were removed with a dedicated program.
Next, we removed hot and bad pixels and divided the 
object frames by the flat frame.
After these processes, we obtained flux images of the 
polarimetric components, 
$F_{0^{\circ}}$ and $F_{90^{\circ}}$,
$F_{90^{\circ}}$ and $F_{0^{\circ}}$,
$F_{45^{\circ}}$ and $F_{135^{\circ}}$,
and $F_{135^{\circ}}$ and $F_{45^{\circ}}$,
separated to the left and right of the image.
The Stokes parameters, $Q$ and $U$, are derived as
\begin{equation}
Q = 
F_{0^{\circ}} - F_{90^{\circ}},
\end{equation}
\begin{equation}
U = 
F_{45^{\circ}} - F_{135^{\circ}}.
\end{equation}
By subtracting the right images from the left images, we obtained
16 images of $Q$ and $-Q$, and 16 images of $U$ and $-U$.
We constructed polarized intensity ($PI$) images by
\begin{equation}
PI = \sqrt{Q^{2}+U^{2}}.
\end{equation}
Assuming that polarization due to interstellar material in front 
of GG Tau is negligible,
the polarized intensity image represents the polarized components
of circumbinary structures.

To derive the polarization degree, $P$, of the circumbinary structures,
the polarized-intensity image needs to be divided by the intensity image,
which contains only the components of the circumbinary structures.
The intensity image of GG Tau (hereafter $I_{\rm tot}$) consists of
the intensity of circumbinary structures as well as that of the central binary.
By subtracting the intensity image
of the central binary ($I_{\ast}$)
from $I_{\rm tot}$, we obtained an intensity image of the circumbinary structure
($I_{\rm disk}$), i.e., $I_{\rm disk} = I_{\rm tot} - I_{\ast}$.
For $I_{\ast}$, we created a pseudo-binary image by duplicating
the images of the PSF reference star.
Detailed description of the PSF subtraction is presented in \cite{Itoh}.
Even with the procedure above, we did not subtract PSF of the GG Tau binary
perfectly.
We attribute this imperfection to
a mismatch between the AO corrections of GG Tau and the PSF
reference star.
As a result, the degrees of polarization have uncertainties as large as 
20 \%.
On the other hand, the angle of polarization and polarized intensity are
trustworthy, 
because these two values were obtained before the PSF subtraction.

\section{Results and Discussion}

\begin{figure}
 \centering
 \includegraphics[width=14.0cm, angle=0]{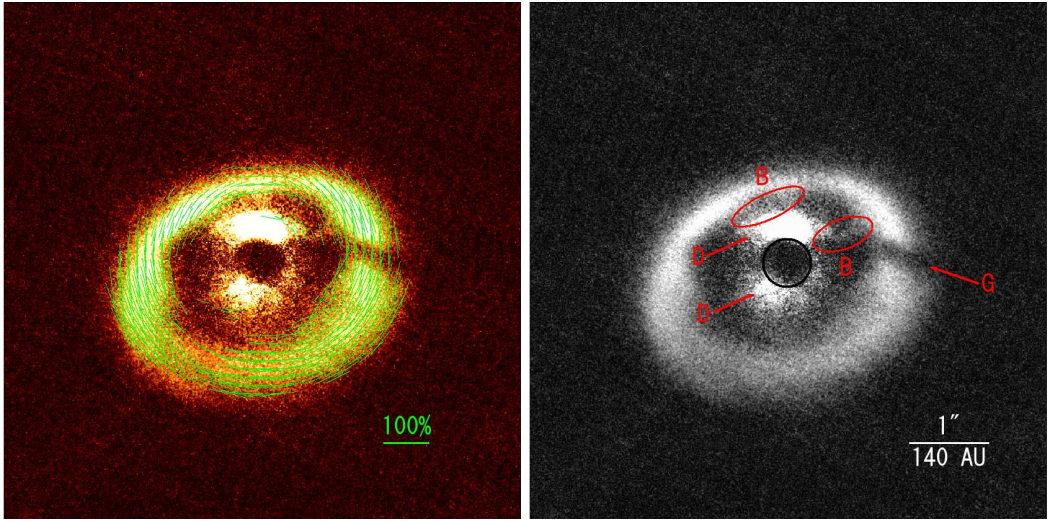}
 \caption{
(left) An $H$-band polarized intensity image of the GG Tau Aa/Ab binary system.
Polarization vectors are overplotted.
Degrees and angles of the polarization in an 11 $\times$ 11 pixel box are 
averaged, if the polarized intensity is detected to more than six $\sigma$
of the sky and if the degree of each pixel is less than 100\%.
The field of view is 6\farcs5$\times$6\farcs5.
North is up and east toward left.
(right) Structures discussed below are indicated on the polarized intensity.
G: the gap structure, D: the circumstellar disks, and B: the bridges.
The black circle indicates the position of the coronagraphic mask.
}
\label{ggpilsc}
\end{figure}

The polarized intensity image of the
GG Tau A binary is shown in Figure \ref{ggpilsc}
with the polarization vectors overlaid.
The central binary was imaged within the coronagraphic mask with a significant
suppression of its light.
The ring-shaped circumbinary disk is clearly seen.
The polarization degrees of the circumbinary disk are between 30 \%
and 100 \%.

%%% gap %%%
The gap structure in the circumbinary disk is evident as a sudden dip to the west.
We propose three hypotheses for this structure:
a less dense region around a planet, 
a shade made by the circumbinary structure,
and a shadow cast by inner material.
Numerical simulations indicate that at an early phase of a planetary formation,
a less dense region appears around a planet (e.g., \citealt{Mayer}).
If the region is optically thin, a gap structure appears in the 
polarized intensity image of the disk.
The less dense region around a planet would orbit the binary with 
the Keplerian velocity.

We detected the orbital motion of the gap structure, by comparing
the image taken in 2001 (\citealt{Itoh}) with that taken in 2011.
Because the inclination of the disk is 37\degr,
the positions of the gap structure were measured in the deprojected disk.
We defined a box as the gap region, 
of which two short sides are interpolations of the
inner and outer edges of the circumbinary disk.
Points at the intersections of these two sides with the middle of the long sides
are defined as the reference points of the gap structure.
The positions
of the reference points are measured from the centroid of the binary.
The mass of each component of the binary is 0.78 M$_{\odot}$ and
0.68 M$_{\odot}$, respectively (\citealt{White}).
The positions of the binary components were accurately measured in the 
2001 image, because the coronagraphic mask has a transmittance of a few
tenths of a percent.
For the 2011 image,
the positions of the binary are measured in the short exposure frames.
The separations between the centroid of the binary and the inner reference point
of the gap structure were 150 AU at 2001 and 160 AU at 2011.
Those of the outer reference point  were 240 AU at 2001 and 260 AU
at 2011.
We consider these discrepancies are due to ambiguous boundaries of the gap region.
The position angle (PA) of the inner reference point was -91\fdg5 at 2001
and -85\fdg6 at 2011.
That of the outer reference point was -101\fdg4 at 2001 and -96\fdg5
at 2011.
Changes in the PAs are +5\fdg9 for the inner reference point and
+4\fdg9 for the outer reference point.
This change corresponds to the orbital periods of 650 yr and 780 yr,
respectively, assuming constant angular velocity.
However, the expected orbital periods are 1700 yr for a point at 160 AU from the
centroid and 3500 yr for a point at 260 AU.
The change in the PA during two observational epochs should be
2\fdg2 and 1\fdg1
for the inner and outer reference points, respectively.
%%%%%%%%%%%%%%%%%%%
Thus, the change of the position angles of the gap is not consistent with the
angular velocity predicted from the Kepler motion of the circumbinary disk.

Moreover, the motion of the gap structure is 
in a direction opposite that of the disk material.
\cite{Kawabe} resolved the circumbinary disk in the millimeter
wavelengths.
The redshifted component of the $^{12}$CO emission is located at
west of the dust continuum and the blueshifted component at east.
\cite{Roddier} observed the GG Tau binary in the near-infrared
wavelengths with the adaptive optics system.
They revealed the orbital motions of the binary components in the
clockwise sense.
Assuming that the orbital angular momenta of the stars and the circumbinary
disk are roughly in the same direction, \cite{Guilloteau}
indicated that the material in the circumbinary disk orbit clockwise
and the disk is inclined such that its northern edge is nearest to us.
On the other hand, comparison of the optical and near-infrared
images of GG Tau reveals that 
the gap structure moves in the opposite direction.
In the $HST$ optical image taken at 1997, the PA of the gap
is $-$92 \degr (\citealt{Krist}).
In the near-infrared image taken at 1998 with $HST$ (\citealt{Silber})
and in that taken at 2001 with the Gemini telescope (\citealt{Potter}) 
the PAs of the gap structure seem around $-90$\degr.
In the near-infrared image taken at 2001, the PA of the gap structure is
$-$91\fdg5 at the inner reference point and $-$101\fdg4 at the outer
reference point (\citealt{Itoh}).
The PAs of the gap structure is $-$85\fdg6 at the inner reference point and
$-$96\fdg5 at the outer reference point at 2011.
These measurements indicate that the motion of the gap structure 
is anti-clockwise and in a direction opposite that of the disk material.
%%%%%%%%%%%%%%%%%%%
%Moreover, \cite{Dutrey} indicated that material in the circumbinary disk
%moves on the Keplerian orbits as the PA decreases.
%The motion of the gap structure is in a direction opposite that of the
%disk material.
We conclude that the gap structure is not a less dense region around a planet 
in the circumbinary disk.

Next, we consider a shade made by the circumbinary structure.
It is expected that the shade appears as a dark region in the polarized
intensity image of the disk, if the disk has a local concave structure.
If the circumbinary disk is a rigid body, a local concave, thus a shade, would
orbit the binary in a prograde direction.
Otherwise, if a retrograde density wave propagates through the circumbinary disk,
a shade moves independently of the disk matter.
However, spiral density waves in a circumstellar disk can be amplified
only if the wave rotates in a prograde sense (e.g. \citealt{Shu}).
We claim that the gap structure is not a shade made by a local concave of 
the circumbinary disk.

\cite{Itoh} proposed that the gap structure may be a shadow of
a material between the central binary and the circumbinary disk.
First, we consider a circumstellar disk of the binary component as an obscuring
structure.
It is known that the circumbinary disk is a thick flared disk.
%%%%%%%%%%%%%%%%%%%%
If the circumstellar disk
is largely inclined to the plane of the circumbinary disk,
illumination of the central star to
the circumbinary disk would be suppressed
along the midplane of the circumstellar disk.
However, \cite{Silber} rejected this hypothesis
due to lack of a second, diametrically opposed gap on the circumbinary disk.
\cite{Krist} proposed a circumstellar disk with
a large azimuthal density enhancement.
Such an enhancement may shadow the circumbinary disk 
to produce the gap structure.
The PA of the gap axis should equal the PA of the gap structure,
if an obscuring structure is spherical or extends only in the radial direction.
%%%%%%%%%%%%%%%%%%%%
%The PA of the gap axis should equal the PA of the gap structure,
%if a spherical clump makes a shadow and the disk is geometrically thin.
However, the PA of the gap axis is  $-$111\degr, which is very different from
the PA of the gap structure.
If the obscuring structure extends in the azimuthal direction,
the structure can create a shadow
whose axis PA is different from the PA of the shadow.
%%%%%%%%%%%%%%%%%%%%
A circumstellar disk with an azimuthal density
enhancement has been reported around several YSOs
(e.g. AB Aur, \citealt{Fukagawa}; V718 Per, \citealt{Grinin}).
The gap structure moves in a retrograde direction.
Assuming that the orbital angular momenta of the stars, the circumstellar disks,
and the circumbinary disk are roughly in the same direction, a
precessing circumstellar disk may account for the shadow.
Such a disk is proposed for the circumbinary disk around an eclipsing young
system, KH 15D (\citealt{Kusakabe}).
As a conclusion, we propose a precessing circumstellar disk with an azimuthal
structure, shadowing a part of the circumbinary disk.

As the obscuring structure other than the circumstellar disk,
%%%%%%%%%%%%%%%%%%%%
\cite{Krist} proposed a dense clump in an accretion stream.
We also imagine that the jet emanating from the secondary star
tilts to the circumbinary disk.
Dust in the jet blocks a part of the light from the primary star,
making a shadow on the circumbinary disk.
To identify the obscuring structure, high-spatial resolution 
observations in close vicinity of the binary system are required.

%%% circumstellar disks %%%
%%% bridge structure %%%
There are emissions between the binary and the circumbinary disk.
Two emissions around the binary extend north and south.
Bridge structures are seen
with the PA between 0\degr~and 40\degr~and with the PA at -60\degr,
respectively.
These structures are not ghosts, because these are polarized.
If they are not polarized, no structure appears in the polarized intensity image.
The polarization vectors depicted at a part of the structures
face the binary, indicating scattering.
We consider that the former and latter structures correspond
to the circumstellar disks and bridges between the circumbinary disk
and the circumstellar disks, respectively.
A bridge structure between a circumbinary disk and circumstellar disks
is suggested in many numerical simulations (e.g. \citealt{Bate};
\citealt{Hanawa}).
Materials in a circumbinary disk are thought to accrete to
circumstellar disks through the bridge structure.
The northern bridge may correspond to the millimeter dust streamer 
(\citealt{Pietu}).
However, because of imperfection of the PSF subtraction,
detections of these structures are marginal.
Further polarimetric observations are required under stable conditions.

\begin{figure}
 \centering
 \includegraphics[width=14.0cm, angle=0]{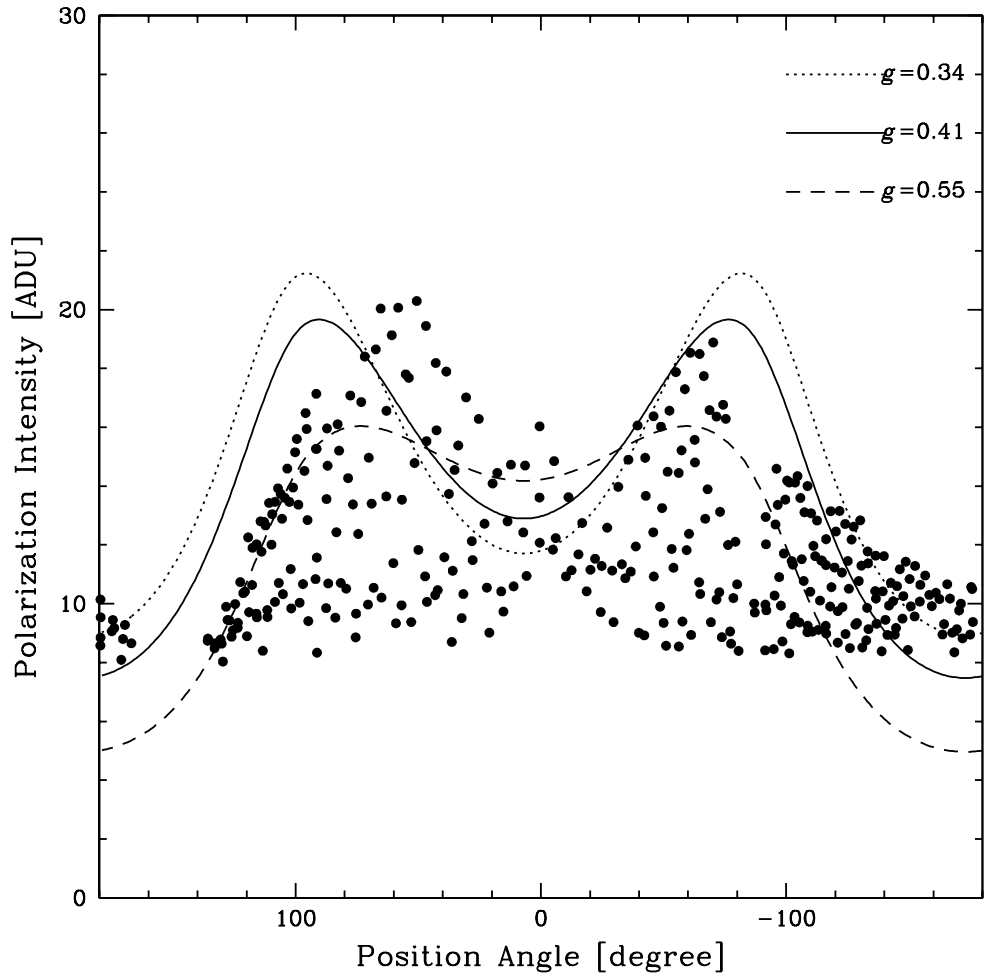}
 \caption{
An azimuthal profile of the polarized intensity of the circumbinary disk
of GG Tau.
The polarized intensities in an 11 $\times$ 11 pixel box are averaged,
if the polarized intensity is detected to more than six $\sigma$
of the sky and if the degree of each pixel is less than 100\%.
With this detection threshold, the polarization vectors cover
the most region of the circumbinary disk (see Fig. 1).
Polarized intensities derived from a simple disk model
are shown by lines.
}
\label{ring_pi}
\end{figure}

Figure \ref{ring_pi} shows the azimuthal profile of the $H$-band polarized intensity
of the circumbinary disk.
In the figure, the polarized intensities derived with a simple
disk model are also shown.
The profile of the polarized intensity is a combination of an
intensity profile and a profile of the polarization degree.
We used the Henyey-Greenstein function and the Rayleigh function 
(\citealt{White79}).
We employed the disk model proposed by \cite{Guilloteau}, i.e.,
the inclination of the disk is 37\degr, the position angle of
the semi-minor axis is 7\degr, and the disk opening angle is 15\degr.
From intensity profiles of the disk in multi-wavelengths,
\cite{Duchene} derived the scattering
$g$ parameter of the Henyey-Greenstein 
function as 0.41 for the best fit and 0.34 -- 0.55 for the acceptable values.
We calculated the polarized intensities with $g$ = 0.34, 0.41, and 0.55.
The polarized intensity will decrease, if the disk surface ripples, for example.
We compared the polarized intensity derived from the disk model with
the maximum of the observed polarized intensity at each position angle.
Among three models, the model with $g$ = 0.41 shows the best fit
to the observed polarized intensity.
Because the disk is optically thick in the $H$-band,
light from the central star is scattered in the surface of the disk.
The observed polarized intensity profile is well reproduced with
Rayleigh-like scattering, indicating small particles in the surface layer.
\cite{Duchene} proposed a stratified structure.
It is claimed that large dust grains settled in the midplane of the disk,
and the disk surface is dominated by small dust grains.
The azimuthal profile of the observed
polarized intensity of the disk is consistent with
this two-layer structure.
Nevertheless,
the discrepancies of the polarized intensity between the observation and the 
model are rather large.
We attribute this mismatch to the use of the simple model.
Comparison of the polarization degree between observations 
and the 3-D disk model with light scattering of non-spherical
dusts will reveal dust-size distribution in the circumbinary disk.

\begin{figure}
 \centering
 \includegraphics[width=14.0cm, angle=0]{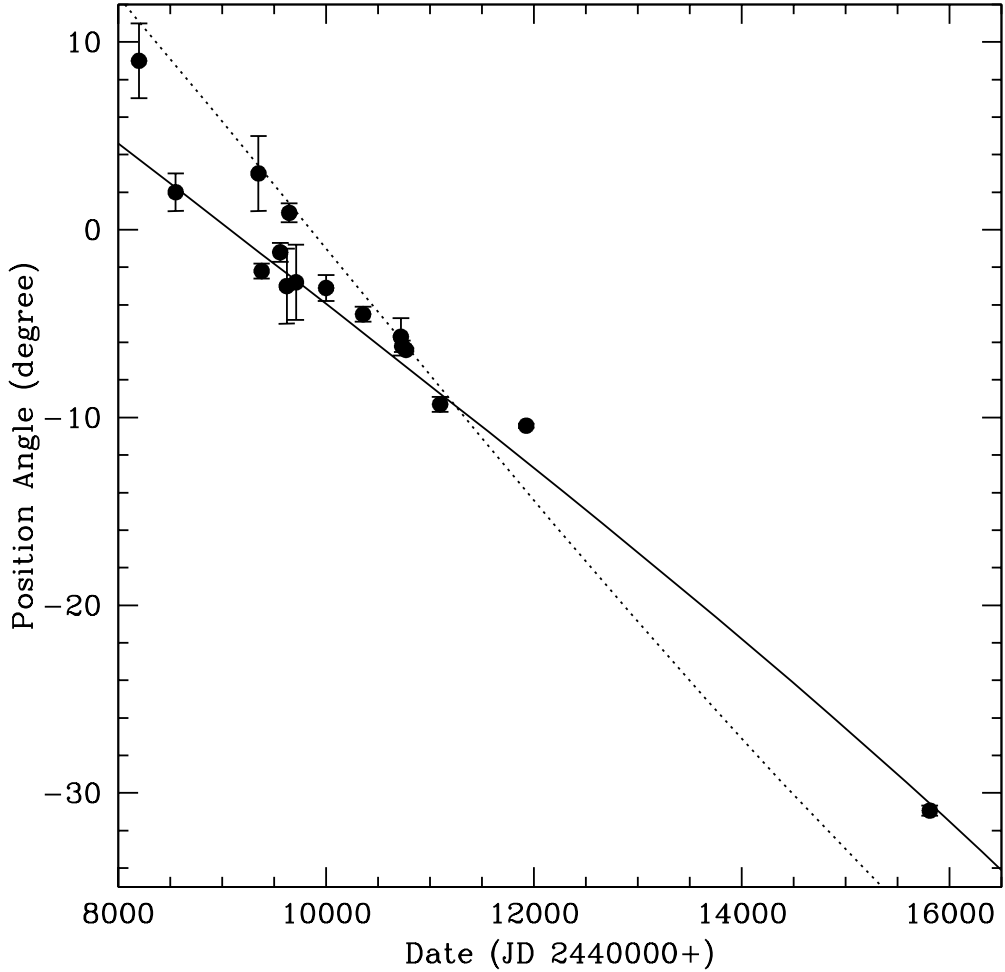}
 \caption{Position angles of the central binary.
The filled circles represent the observed position angles.
The solid line indicates the position angles for the orbit with $a$ = 33.4 AU
and $e$ = 0.34.
The dotted line indicates the position angles for the orbit with $a$ = 62 AU 
and $e$ = 0.35.
}
\label{orbitpa}
\end{figure}

%%% binary %%%
The orbital motion of the central binary was also examined.
The separation and position angle of the binary at 2011
are 256.0$\pm$1.6 mas and 
-30\fdg9$\pm$0\fdg3 in the projected plane.
The separation has been almost constant for 20 years.
Figure \ref{orbitpa} shows the position angles of the binary in the last
20 years.
\cite{Beust2005} proposed two orbits for the binary.
The small orbit has $a = 33.4$ AU and $e = 0.34$.
The large orbit has $a = 62$ AU and $e = 0.35$.
The expected position angles for these two orbits are also shown in Figure 
\ref{orbitpa}.
For the large orbit, the
projected separation is as large as 288 mas even at the periastron.
For this orbit, we draw the position angles such that the
companion passed the periastron at JD = 2,450,000 (circa 1995).
The position angle observed in 2011 is consistent with the small orbit,
but is not consistent with the large orbit.
\cite{Beust2005} and \cite{Beust2006} pointed out that the inner edge of 
the circumbinary disk is approximately twice as large as should
be expected with the small binary orbit.
As a cause of this mismatch, 
they proposed a secular evolution of the GG Tau A orbit and a massive
circumbinary planet around $\sim$ 140 AU from the GG Tau A binary.

\section{Conclusions}
A high angular resolution near-infrared polarized-intensity image of the GG
Tau A binary system was obtained with the Subaru Telescope.
\begin{enumerate}
\item The image shows the circumbinary disk scattering the light from 
the central binary.
Combined with a previous observation of the circumbinary disk,
our image indicates that the gap structure in the circumbinary disk orbits
anti-clockwise.
On the other hand, material in the disk orbit clockwise.
We conclude that the gap structure is not a less dense region around a circumbinary
planet, nor a shade made by a local concave of the disk, but
a shadow of a material located between the binary and the circumbinary disk.
\item The azimuthal profile of the polarized intensity of the circumbinary disk 
is roughly reproduced by a simple disk model
with the Henyey-Greenstein function and the Rayleigh function, 
indicating small dust grains
at the surface of the disk.
\item The separations and the position angles of the stellar components of the
binary in the past 20 years are consistent with the binary orbit with
$a = 33.4$ AU and $e = 0.34$.
\end{enumerate}

\begin{acknowledgements}
We thank Dr. Michihiro Takami for useful discussion.
Y. I. is supported by a Grant-in-Aid for Scientific Research No. 24540231.
J. C. is supported by the U.S. National Science Foundation under Award 
No. 1009203.
\end{acknowledgements}

\clearpage

\bibliographystyle{raa}
\bibliography{ggtau}

\end{document}